\documentclass[twocolumn,showpacs,preprintnumbers,amsmath,amssymb, showkeys]{revtex4}

\usepackage{graphicx}
\usepackage{dcolumn}
\usepackage{bm}
\usepackage{epsf}

\begin{document}


\title{Spatial effects of Fano resonance in local tunneling conductivity in vicinity of impurity on semiconductor surface}

\author{V.\,N.\,Mantsevich}
 \altaffiliation{vmantsev@spmlab.phys.msu.ru}
\author{N.\,S.\,Maslova}%
 \email{spm@spmlab.phys.msu.ru}
\affiliation{%
 Moscow State University, Department of  Physics,
119991 Moscow, Russia
}%

\date{\today }
4 pages, 3 figures
\begin{abstract}
We present the results of local tunneling conductivity spatial
distribution detailed theoretical investigations  in vicinity of
impurity atom for a wide range of applied bias voltage. We observed
Fano resonance in tunneling conductivity resulting from interference
between resonant tunneling channel through impurity energy level and
direct tunneling channel between the tunneling contact leads. We
have found that interference between tunneling channels strongly
modifies form of tunneling conductivity measured by the scanning
tunneling microscopy/spectroscopy (STM/STS) depending on the
distance value from the impurity.
\end{abstract}

\maketitle

\section{Introduction}

    Localized states of individual impurity atoms play the key role in
tunneling processes in small size junctions and often determine the
behavior of tunneling characteristics in STM/STS contacts.
Investigation of tunneling through impurity atom energy level reveal
interesting phenomena such as resonant tunneling, interference
between resonant and direct tunneling channels, Kondo effects in
quantum dots strongly coupled with the leads of tunneling contact
\cite{Konig} and Fano-type line shapes in the situation of
multichannel transport \cite{Fano,Gores}. Influence of different
impurities on the tunneling conductivity was widely studied
experimentally and theoretically. Most of the experiments were
carried out  with the help of scanning tunneling
microscopy/spectroscopy technique \cite
{Dombrowski,Boon,Maslova,Panov}, but experimental measurements
provides no information weather electron transport occurs coherently
or incoherently. Answer to this question can be found from the
comparison between experimental and theoretical results. Theoretical
investigations of single impurities influence on tunneling
conductivity mostly deals with Green's functions formalism \cite
{Hofstetter, Konig}.

Usually theoretical calculations of interference effects between
resonant and direct tunneling  channels correspond to the case when
metallic STM tip is positioned above the impurity atom \cite
{Hofstetter, Konig}. However experimentally tunneling conductivity
can be measured far away from the impurity \cite{Madhavan} and just
this case is quite of great interest because it gives an opportunity
to initialize impurities types. So in our work we present formula,
which describe spatial distribution of local tunneling conductivity
in vicinity of impurity in the case of interference between resonant
and direct tunneling channels. We found that taking into account
real part of impurity Green's function leads to drastical changing
of the tunneling conductivity form depending on the values of
tunneling rates and distance.
\section{The suggested model and main results}
\begin{figure}[h]
\centering
\includegraphics[width=70mm]{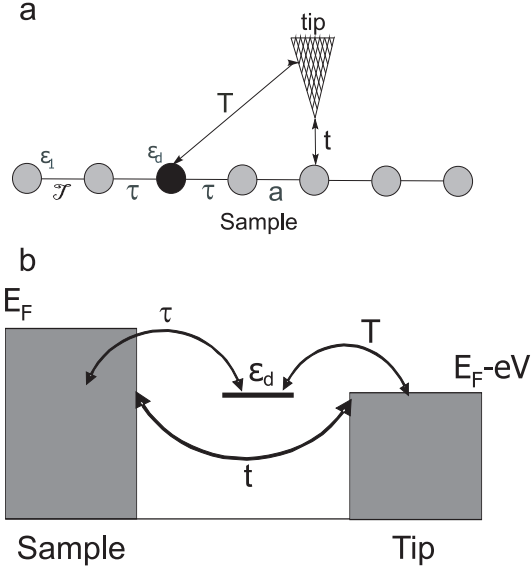}%
\caption{a) Schematic view of experimental geometry; b)Schematic
diagram of direct and resonant tunneling channels.} \label{1}
\end{figure}

 We shall analyze tunneling between semiconductor surface ($1D$ atomic chain)
and metallic STM tip (Fig. \ref{1}a). $1D$ atomic chain is formed by
the similar atoms with energy levels $\varepsilon_{1}$ and similar
tunneling transfer amplitudes $\Im$ between the atoms along the
atomic chain. Distance between the atoms in the atomic chain is
equal to $a$. Atomic chain includes impurity with energy
$\varepsilon_{d}$, tunneling transfer amplitude from impurity atom
to the nearest atoms of the atomic chain is $\tau$. Tunneling
conductivity is measured by the STM tip. Direct tunneling between
the surface continuous spectrum states and tip states is described
by the transfer amplitude $t$. Tunneling from the impurity energy
level to the tip is described by tunneling transfer amplitude $T$.
The model of tunneling contact formed by semiconductor and metallic
tip is depicted in Fig. \ref{1}b.

The model system can be described by the Hamiltonian: $\Hat{H}$:

$$\Hat{H}=\Hat{H}_{0}+\Hat{H}_{imp}+\Hat{H}_{tun}+\Hat{H}_{tip}$$
\begin{eqnarray}
&\Hat{H}_{0}&=\sum_{k}\varepsilon_{k}c_{k}^{+}c_{k}+\sum_{k}\Im c_{k}^{+}c_{k}+h.c.\nonumber\\
&\Hat{H}_{tun}&=\sum_{k,d}\tau c_{k}^{+}c_{d}+\sum_{d,p}T c_{d}^{+}c_{p}+\sum_{k,p}tc_{k}^{+}c_{p}+h.c.\nonumber\\
&\Hat{H}_{imp}&=\sum_{d}\varepsilon_{d}c_{d}^{+}c_{d};\Hat{H}_{tip}=\sum_{p}\varepsilon_{p}c_{p}^{+}c_{p}\
\end{eqnarray}
Indices $k$ and $p$ label the states in the left (semiconductor) and
right (tip) lead, respectively. The index $d$ indicates that
impurity electron operator is involved. $\Hat{H}_{0}$ is a typical
Hamiltonian for atomic chain with hoppings without any impurities.
$\Hat{H}_{tun}$ describes resonant tunneling transitions from the
impurity state to the semiconductor and metallic tip and direct
transitions between the tunneling contact leads. $\Hat{H}_{imp}$
corresponds to the electrons in the localized state formed by the
impurity atom in the atomic chain, $\Hat{H}_{tip}$ describes
conduction electrons in the metallic tip.

    We shall use diagram technique in our
investigation of tunneling conductivity spatial distribution
\cite{Keldysh}. Using Keldysh functions $G^{<}$ the tunneling
current from the right lead can be determined as (we set charge
$e=1$):

\begin{eqnarray}
I(V)&=&Im(J(V))\nonumber\\
J(V)&=&i\sum_{k,p}\int d\omega(TG^{<}_{pd}+tG^{<}_{pk})+h.c.\ \label
{current}
\end{eqnarray}
where $J(V)$ is tunneling "response function".

    The first (second) part of the Eq. \ref{current} describes electron
transfer from the right lead to the impurity (to the left lead) or
vice versa. Our aim is to re-write Keldysh functions $G^{<}_{pd}$
and $G^{<}_{pk}$ in the terms of retarded $G^{R}_{pd(pk)}$ and
advanced $G^{A}_{pd(pk)}$ Green's functions. We can do it with the
help of Dyson-like equations
\begin{eqnarray}
G^{<}_{pd}&=&(G^{0}_{pp}TG_{dd})^{<}+(G^{0}_{pp}tG_{kd})^{<}\nonumber\\
G^{<}_{dd}&=&G^{0<}_{dd}+(G^{0}_{dd}\tau
G_{kd})^{<}+(G^{0}_{dd}TG_{pd})^{<}\nonumber\\
G^{<}_{kd}&=&(G^{0}_{kk}\tau G_{dd})^{<}+(G^{0}_{kk}tG_{pd})^{<}\
\label{system}
\end{eqnarray}
for the function $G^{<}_{pd}$ and equations
\begin{eqnarray}
G^{<}_{pk^{'}}&=&\sum_{k^{''}}(G^{0}_{pp}tG_{k^{''}k^{'}})^{<}+(G^{0}_{pp}TG_{dk^{'}})^{<}\nonumber\\
G^{<}_{dk^{'}}&=&\sum_{k^{''}}(G^{0}_{dd}\tau
G_{k^{''}k^{'}})^{<}+(G^{0}_{dd}TG_{pk^{'}})^{<}\nonumber\\
G^{<}_{k^{''}k^{'}}&=&G^{0<}_{kk}+\sum_{p}(G^{0}_{kk}t
G_{pk^{'}})^{<}+(G^{0}_{kk}\tau G_{dk^{'}})^{<}\ \label{system1}
\end{eqnarray}
for the function $G^{<}_{pk^{'}}$.

To evaluate the dependence of local tunneling conductivity on the
distance from the impurity one have to write instead of tunneling
transfer amplitude $t$ the expression $te^{ikx}e^{-ipx}$ which takes
into account spatial distribution of tunneling conductivity
(x-distance from the impurity along the atomic chain). Substituting
the correspondent expressions for the Keldysh functions, received
from equations \ref{system}, \ref{system1} to the tunneling current
equation \ref{current}, performing summation over $k^{}$ and $k^{'}$
and taking imaginary part we get the expression for the local
tunneling conductivity:

\begin{eqnarray}
\frac{dI}{dV}&=&\sqrt{\gamma_{kp}\gamma_{kd}\gamma_{pd}\nu^{0}_{k}}ReG^{R}_{dd}cos(2k(\omega)x)+\nonumber\\
&+&\gamma_{kp}(\gamma_{kd}+\gamma_{pd})\nu^{0}_{k}ImG^{R}_{dd}cos(2k(\omega)x)+\nonumber\\
&+&\frac{\gamma_{kd}\gamma_{pd}}{\gamma_{kd}+\gamma_{pd}}ImG^{R}_{dd}+\frac{\gamma_{kd}^{2}\gamma_{pd}\gamma_{kp}(\omega-\varepsilon_{d})cos(2k(\omega)x)}{((\omega-\varepsilon_{d})^{2}+(\gamma_{kd}+\gamma_{pd})^{2})^{2}}+\nonumber\\
&+&\gamma_{kp}\nu^{0}_{k}(1+\frac{(\omega-\varepsilon_{d})^{2}+\gamma_{kd}^{2}-\gamma_{pd}(\omega-\varepsilon_{d})}{(\omega-\varepsilon_{d})^{2}+(\gamma_{kd}+\gamma_{pd})^{2}})\cdot\nonumber\\
&\cdot&(\frac{(\omega-\varepsilon_{d})^{2}+(\gamma_{kd}+\gamma_{pd})^{2}(1-cos(2k(\omega)x))}{(\omega-\varepsilon_{d})^{2}+(\gamma_{kd}+\gamma_{pd})^{2}}+\nonumber\\
&+&\frac{\gamma_{pd}(\gamma_{pd}+\gamma_{kd})cos(2k(\omega)x)}{(\omega-\varepsilon_{d})^{2}+(\gamma_{kd}+\gamma_{pd})^{2}})\
\end{eqnarray}
where impurity retarded Green's function is defined by the
expression:
\begin{eqnarray}
G^{R}_{dd}=\frac{1}{\omega-\varepsilon_{d}-i(\gamma_{kd}+\gamma_{pd})}
\end{eqnarray}
Relaxation rates $\gamma_{kd}$, $\gamma_{pd}$  are determined by
electron tunneling transitions from localized states to the leads
$k$ and $p$ continuum states and relaxation rate $\gamma_{kp}$
corresponds to direct tunneling transitions between $k$ and $p$
continuum states $\sum_{p}T^{2}ImG_{pp}^{0R}=\gamma_{pd}$;
$\sum_{p}t^{2}ImG_{pp}^{0R}=\gamma_{kp}$;
$\sum_{k}\tau^{2}ImG_{kk}^{0R}=\gamma_{kd}$.

$\nu^{0}_{k}$ is unperturbed density of states in semiconductor.
Expression for $k_{x}(\omega)$ can be found from the dispersion law
of the $1D$ atomic chain which has the form
\begin{eqnarray}
\omega(k_{x})=2\Im\cdot\cos(k_{x}a)\
\end{eqnarray}

\begin{figure*}
\centering
\includegraphics[width=160mm]{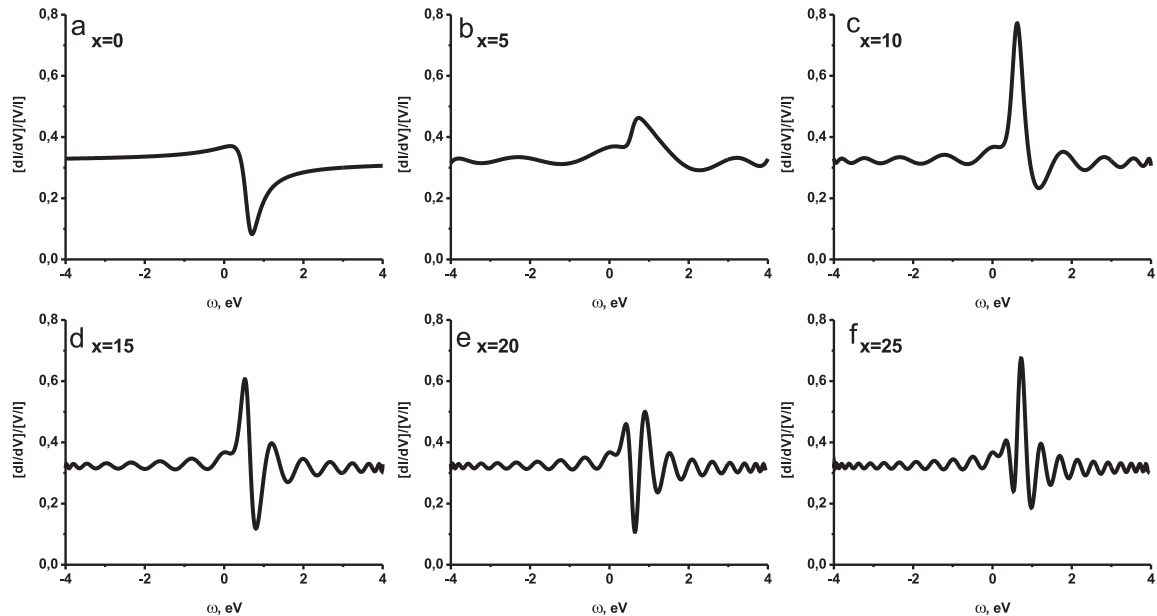}%
\caption{Local tunneling conductivity as a function of applied bias
voltage calculated for different values of the distance $x$ from the
impurity atom along the atomic chain. For all the figures values of
the parameters $a=1$, $t_{kp}=0,40$, $T_{pd}=0,25$, $\tau=0,35$,
$\Im=1,00$, $\varepsilon_{d}=0,60$, $\nu^{0}_{k}=1$ are the same.}
\label{2}
\end{figure*}

\begin{figure*}[t]
\centering
\includegraphics[width=160mm]{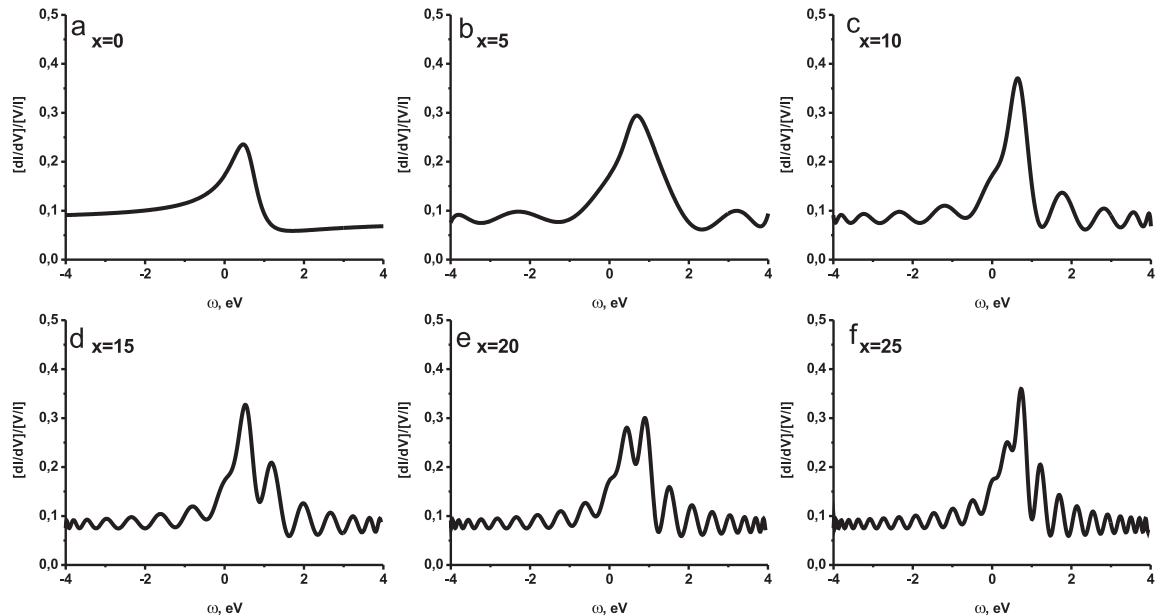}%
\caption{Local tunneling conductivity as a function of applied bias
voltage calculated for different values of the distance $x$ from the
impurity atom along the atomic chain. For all the figures values of
the parameters $a=1$, $t_{kp}=0,20$, $T_{pd}=0,40$, $\tau=0,50$,
$\Im=1,00$, $\varepsilon_{d}=0,60$, $\nu^{0}_{k}=1$ are the same.}
\label{3}
\end{figure*}

Figures \ref{2} and \ref{3} show tunneling conductance as a function
of applied bias voltage for different values of distance from the
impurity and different values of tunneling transfer amplitudes.
Figure \ref{2} corresponds to the case when value of direct
tunneling channel transfer amplitude $t$ exceeds values of resonant
tunneling channel transfer amplitudes through the impurity energy
level $T$ and $\tau$. In this case tunneling conductivity calculated
above the impurity (Fig. \ref{2}a) has Fano line shape due to
interference between resonant and nonresonant tunneling channels.
Tunneling conductivity shows a dip when applied bias voltage is
equal to the impurity energy level position
$(\omega=\varepsilon_{d})$. The degree of asymmetry of the resonant
line shape depend on the relative strengths of transmission through
the two channels. When the distance value is not equal to zero a
series of dips in the local tunneling conductivity exist. Amount of
dips increases with the increasing of distance value. When applied
bias voltage is equal to the impurity energy level position
(resonance) not only a dip (Fig.~\ref{2}a,e) but also a peak
(Fig.~\ref{2}b,f) can exist in the tunneling conductivity measured
aside from the impurity. At the fixed parameters of the atomic chain
such as tunneling transfer amplitudes existance of a dip or a peak
in tunneling conductivity in the resonance is determined by the
value of the distance. At the special values of distance it can be
neither resonance dip nor resonance peak (Fig.~\ref{2}d).
Replacement of the dip by the peak in tunneling conductivity when
applied bias voltage is equal to the impurity energy level position
is the result of interference between the direct and resonant
tunneling channels (between real and imaginary parts of impurity
Green's functions).

    Figure \ref{3} shows tunneling conductance as a
function of applied bias voltage in the case when values of resonant
tunneling channel transfer amplitudes through the impurity energy
level $T$ and $\tau$ exceed  transfer amplitude of direct tunneling
channel $t$. In this case formation of a asymmetric peak in
tunneling conductivity calculated above the impurity (Fig. \ref{3}a)
in the resonance when applied bias voltage is equal to the impurity
energy level position $(\omega=\varepsilon_{d})$ can be seen in the
tunneling conductivity. With increasing of distance value resonant
peak still exists but it's shape changes due to the effects caused
by the interference between tunneling channels (Fig. \ref{3}b,c).
Simultaneously formation of non-resonance dips in local tunneling
conductivity is found with the increasing of distance.

    Further increasing of distance value leads to formation of the
resonance dip when applied bias voltage is equal to the impurity
energy level position (Fig. \ref{3}d-f). From the other hand dip's
formation can be interpreted as resonance peak splitting. Dip's
amplitude depends on the distance value and it decreases with the
distance increasing at the fixed values of tunneling rates in the
system. More over with increasing of the distance interference
between the direct and resonant tunneling channels influence the
amplitudes of the peaks formed by resonant peak splitting. Amplitude
of the peak which corresponds to the lower value of applied bias
voltage decreases with increasing of the distance value from the
impurity. Amplitude of the peak which corresponds to the higher
value of applied bias voltage increases with increasing of the
distance value from the impurity Fig. \ref{3}d-f).

\section{Conclusion}
 In this work we have studied the influence of interplay between the
resonant tunneling channel through the impurity energy level and
direct tunneling channel between the leads of tunneling contact on
local tunneling conductivity. It was found that depending on the
value of the distance from the impurity special features can be seen
in tunneling conductivity, such as dip or peak when applied bias
voltage is equal to the impurity energy level position. When value
of direct tunneling channel transfer amplitude exceeds values of
resonant tunneling channel transfer amplitudes through the impurity
energy level tunneling conductivity calculated above the impurity
has Fano line shape due to interference between resonant and
nonresonant tunneling channels. With increasing of the distance from
the impurity a series of dips are formed in local tunneling
conductivity in spite of a dip or a peak exists in the resonance. We
derived replacement of the dip by the peak as a result of
interference between the direct and resonant tunneling channels.

Replacement of the peak by the dip and vice versa can be observed
experimentally with the help of STM/STS technique as impurity atom
switching on and off in local tunneling conductivity.

This work was  supported by RFBR grants and by the National Grants
for technical regulation and metrology.


\pagebreak

\end{document}